\documentclass[amsmath,12pt,amssymb,preprint,nofootinbib,floatfix]{revtex4}
\usepackage{amsfonts}
\usepackage{lipsum}
\allowdisplaybreaks
\usepackage{graphicx} 
\usepackage{epsfig}
\usepackage{bm}
\usepackage[symbol]{footmisc}
\usepackage{xcolor}

\begin{document}

\title{Open Charm Mesons and Charmonium states in Magnetized Strange Hadronic Medium at Finite Temperature}
\author{Amal Jahan C.S.}
\email{amaljahan@gmail.com}
\author{Amruta Mishra}
\email{amruta@physics.iitd.ac.in}
\affiliation{Department of Physics, Indian Institute of Technology,Delhi, Hauz Khas, New Delhi - 110016, India}

\begin{abstract}
 
 We investigate the masses of the pseudoscalar ($D$($D^0$, $D^+$), $\bar{D}$($\bar{D^0}$, $D^-$) and vector open charm mesons ($D^*$($D^{*0}$, $D^{*+}$), ${\bar{D}}^*$(${\bar{D}}^{*0}$, $D^{*-}$) as well as the pseudoscalar ($\eta_c(1S)$, $\eta_c(2S)$) and the vector charmonium states ($J/\psi$, $\psi(2S)$, $\psi(1D)$) in the asymmetric hot strange hadronic medium in the presence of strong magnetic fields. In the magnetized medium, the mass modification of open charm mesons due to their interactions with baryons and the scalar fields ($\sigma$, $\zeta$, and  $\delta$) are investigated in a chiral effective model. Moreover, the charged pseudoscalar meson ($D^\pm$), as well as the longitudinal component of charged vector meson ($D^{*\pm \parallel}$), experience additional positive mass modifications in the magnetic field due to Landau quantization. The effect of the modification of gluon condensates simulated by the medium change of dilaton field $\chi$ on the masses of the charmonia is also calculated in the chiral effective model. The contribution of masses of light quarks is also considered in the modification of gluon condensates. At high temperatures, the magnetically induced modifications of scalar fields significantly reduce the in-medium masses of mesons. The effects of magnetically induced spin mixing between the pseudoscalar and the corresponding vector mesons are incorporated in our study. The spin-magnetic field interaction of these mesons is considered through a phenomenological effective Lagrangian interaction. The spin mixing result in a positive mass shift for the longitudinal component of the vector mesons and a negative mass shift for the pseudoscalar mesons in the presence of the magnetic field. From the obtained in-medium mass shifts of charmonia and open charm mesons, we have also calculated the partial decay widths of $\psi(1D)$ to $D\bar{D}$, using a light quark pair creation model, namely the $^3P_0$ model. Spin mixing and strangeness fraction enhance the partial decay width at small magnetic fields.
\end{abstract}
\maketitle
\vspace{-1.0cm}

\section{INTRODUCTION}

The effect of strong magnetic fields on the properties of hadrons has recently received significant interest due to the phenomenological consequences in the relativistic heavy-ion collision experiments. The strength of the magnetic fields in such experiments could be as enormous as $eB \sim 2{m_{\pi}}^2$ $\sim 6\times 10^{18}$ Gauss in the Relativistic Heavy Ion Collider(RHIC) at Brookhaven National Laboratory (BNL) and eB $\sim 15{m_{\pi}}^2$ $\sim 10^{19}$  Gauss in the Large Hadron Collider(LHC) at CERN \cite{kharzeev,Fukushima,skokov, Dengand}. These studies indicate that the magnitude of the magnetic field depends on the energy of the collision, as well as the impact parameter, and is produced at the early stages of collision. Since charm quarks, due to their large mass are  formed by initial hard scatterings at the early stages of heavy ion collision, the charm quark systems are sensitive to magnetic fields \cite{Cho, Cho1}. Hence the properties of charmonium states and open charm mesons will undergo modifications in magnetic fields and must be investigated. 

Strong magnetic fields can modify the internal structure of mesons as well as the QCD condensates \cite{Cho,Cho1}. The chiral condensates in QCD, which is a measure of the spontaneous chiral symmetry breaking of the system, is modified by the magnetic field  through the phenomena of magnetic catalysis at low temperature and inverse magnetic catalysis at very high temperature  \cite{kharzeev_springer}. The gluon condensates, results in the scale invariance breaking or scale anomaly by which the trace of enery-momentum tensor becomes non-zero. The gluon condensates are also modified in strong magnetic fields through gluon catalysis. \cite{gluoncatalysis1,gluoncatalysis2,gluoncatalysis3,gluoncatalysis4}. The condensates also modify with variation in medium conditions such as baryon density, temperature, isospin asymmetry and the strangeness content of the medium. When the hadrons interact with these condensates, their properties such as masses and decay widths also modify in magnetic field at finite baryon density and temperature. 

Open heavy flavor mesons and various heavy quarkonium states in magnetic fields have been investigated in QCD sum rule approach \cite{Cho, Cho1, Machado, Gubler, AM_sumrule_spinmixing, bottomonia_sumrule_Pallabi,catalysis_new1, Rajeshkumar, Rajeshkumar1, Rajeshkumar2} as well as using potential models \cite{Alford, Bonati, Suzuki, Yoshida, Machado2}. The masses of heavy flavor mesons are investigated in the chiral effective model in cold magnetized nuclear medium \cite{SReddy, Dhale, Amal1, Amal2_upsilon} and in magnetized strange hadronic medium \cite{Magstrange, Hotmagstrange}. The partial decay widths of charmonium states to $D\bar{D}$ in the cold magnetized nuclear medium are studied in Refs. \cite{AM_charmdecaywidths_mag, AM_SPM3}. Besides the modification of condensates, the magnetic field also introduces the mixing of the spin eigenstates between the spin-singlet states and spin-triplet states of heavy flavor mesons \cite{Cho, Cho1, Gubler, AM_sumrule_spinmixing, bottomonia_sumrule_Pallabi, Alford, Bonati, Suzuki, Yoshida, Machado2, AM_SPM4, AM_SPM5, AM_SPM6,catalysis_new1,catalysis_new2,catalysis_new3,catalysis_new4,catalysis_new5,catalysis_new6}. Under strong magnetic fields, a part of spatial rotation symmetry is broken, and only the azimuthal component along the direction of the magnetic field remains \cite{Gubler}. Consequently, the spin state, which can act as a good quantum number for the meson, is the one along the direction of the magnetic field. Hence the pseudoscalar meson mix with the longitudinal component of the corresponding vector meson. The transverse component of the vector charmonium state does not take part in this mixing. In Ref.\cite{Yoshida}, along with the spin mixing effect, the Zeeman splitting between the transverse components of the spin-triplet states of open charm mesons are also investigated. 

In Ref.\cite{AM_SPM4}, the vacuum masses of the pseudoscalar and vector charm mesons in the presence of the magnetic field are studied with the spin mixing effect incorporated through a phenomenological Lagrangian interaction. The masses of charmonia in the cold magnetized nuclear medium have been calculated, accounting for their spin mixing effect in Ref.\cite{AM_SPM5}, and the same investigation for open bottom mesons and bottomonium states is carried out in Ref.\cite{AM_SPM6}. In Ref.\cite{AM_SPM5}, the partial decay width of $\psi(1D)$ to $D\bar{D}$  in the cold magnetized nuclear medium has been calculated using the field-theoretic model for composite hadrons as well as using an effective hadronic model. In this study the effect of spin mixing effect on the in-medium masses of open charm mesons was not considered . Recently the properties of heavy flavor mesons in the cold magnetized nuclear medium have been investigated incorporating the effect of magnetic catalysis and spin mixing effect \cite{catalysis_new1,catalysis_new2,catalysis_new3,catalysis_new4,catalysis_new5,catalysis_new6}.

In Ref.\cite{Magstrange}, we have investigated the masses of pseudoscalar open charm mesons and vector charmonia in the magnetized strange medium  without considering the effects of spin mixing and finite temperature. In the present work, we have investigated the masses of the pseudoscalar and the vector open charm mesons, as well as the pseudoscalar and the vector charmonium states in the hot magnetized strange hadronic medium. In the present study, the magnetically induced mixing of the pseudoscalar and vector mesons is also taken into account through the effective Lagrangian interaction \cite{Cho, Cho1, Gubler, AM_SPM4, AM_SPM5}. In the magnetized medium, the mass modifications of the pseudoscalar open charm mesons accounting for the modification of light quark condensates and the mass modifications of charmonium states accounting for the modification of gluon condensates are calculated within the chiral effective model \cite{SReddy, Amal1, Magstrange}. Within the chiral model, the modifications of light quark condensates are calculated from the modification of the scalar fields ($\sigma$, $\zeta$, $\delta$), and that of gluon condensates are calculated from the medium change of the dilaton field ($\chi$), introduced through a scale breaking term in the Lagrangian. The contribution of the mass term of light quarks to the modification of gluon condensates is also incorporated in the present study \cite {Magstrange}. The in-medium masses of the vector $D^*$,${\bar{D}}^*$ mesons are calculated by assuming that the magnitude of their mass shifts due to the interaction with nucleons and scalar fields are similar to that of pseudoscalar open charm mesons. The charged open charm mesons, $D^\pm$($D^{*\pm \parallel}$), experience additional positive mass shift in magnetic fields due to Landau quantization. From the mass modifications of both $\psi(1D)$ and open charm mesons accounting for the spin mixing effect, we compute the decay widths of $\psi(1D)$ to $D\bar D$ pair in the magnetized medium using a light quark pair creation model called the $^3P_0$ model \cite{Friman, AM_charmdecaywidths_mag}.  

The outline of the paper is as follows. In section II, we describe the effect of magnetized medium on the mass modifications of open charm mesons and charmonium states using a chiral effective model. We also describe the incorporation of magnetically induced spin mixing on the masses of pseudoscalar and vector mesons using a phenomenological interaction Lagrangian in the same section. Section III describes the mathematical formalism of the $^3P_0$ model and the expressions for the partial decay widths of $\psi(1D)$ to $D\bar{D}$ pair. In Section IV, we discuss and analyze the results obtained and later summarize our findings in section V.

\section{MASSES OF OPEN CHARM MESONS AND CHARMONIA IN HOT MAGNETIZED MEDIUM}

In this section, we discuss the modifications of the masses of pseudoscalar (P) and vector (V) open charm mesons and charmonium states in  strange hadronic matter at finite temperatures in strong magnetic fields. In magnetized medium,  the open charm mesons will have mass shifts due to the medium modification of light quark condensates, whereas charmonium states experience mass modifications due to the modification of gluon condensates. Such modifications are taken into consideration using a chiral effective model. The Hadronic Lagrangian density in chiral effective model \cite{SReddy, AM_SPM5, Magstrange} is given as
\begin{eqnarray}
\mathcal{L_{\textrm{eff}}} = \mathcal{L_\textrm{kin}} + \sum_{W=X,Y,A,V,u}{\mathcal{L_\textrm{BW}}} + \mathcal{L_\textrm{vec}} + \mathcal{L_\textrm{0}}+ \mathcal{L_\textrm{scale break}} + \mathcal{L_\textrm{SB}} + \mathcal{L^{\textrm{B}\gamma}_\textrm{mag}}.
\label{genlag}
\end{eqnarray}

In this equation, $\mathcal{L_\textrm{kin}}$ refers to the kinetic energy terms of the mesons and baryons. $\mathcal{L_\textrm{BW}}$ is the baryon-meson interaction term, where the index $\mathcal{\textrm{W}}$ covers both spin-0 and spin-1 mesons. $\mathcal{L_\textrm{vec}}$ concerns the dynamical mass generation of the vector mesons through couplings with scalar mesons, apart from bearing the quartic self-interaction terms of these mesons. $\mathcal{L_\textrm{0}}$ contains the meson-meson interaction terms introducing the spontaneous breaking of chiral symmetry, and $\mathcal{L_\textrm{scale break}}$ incorporates the scale invariance breaking of QCD through a logarithmic potential given in terms of scalar dilaton field $\chi$. $\mathcal{L_\textrm{SB}}$ corresponds to the explicit chiral symmetry breaking term. Finally $\mathcal{L^{\textrm{B}\gamma}_\textrm{mag}}$ is the contribution by the magnetic field which describe the interactions of octet baryons with the magnetic field. We choose the magnetic field to be uniform and along
the z-axis. The field term of the magnetic part of the Lagrangian has vectorial as well as tensorial interactions with the electromagnetic field \cite{SReddy}. The tensorial interaction is related to the anomalous magnetic moment (AMM) of the baryons \cite{SReddy, Magstrange} .

We then invoke mean-field approximation, where fermions are treated as quantum fields and mesons as classical fields. In this approximation, only the scalar and the vector fields contribute as the expectation values vanish for all other terms. The magnetic field introduces Landau quantization for the charged baryons. For neutral baryons, the magnetic field's effect is only through AMM effects. The effects of temperature can be introduced through Fermi distribution functions in the expressions of scalar and number densities of baryons \cite{Hotmagstrange}. The scalar fields depend on the scalar density of baryons and will be modified with changes in baryon density, magnetic field, strangeness content of the medium, and temperature \cite{Magstrange, Hotmagstrange}. The equations of motion for scalar fields (the non-strange field $\sigma$, strange field $\zeta$, isovector field $\delta$, and the dilaton field $\chi$) are solved as functions of the magnetic field for isospin asymmetric hadronic matter ($\eta=0.5$) at nuclear matter saturation density for different values of strangeness fraction ($f_s$) and temperature ($T$). The modifications of scalar fields in the hot magnetized strange hadronic medium in the chiral effective model are already investigated in Ref.\cite{Hotmagstrange}. 

The in-medium masses of open charm mesons are investigated using the chiral effective Lagrangian approach. Here the chiral $SU(3)$ model has been generalized to chiral $SU(4)$ to include the charm degrees of freedom \cite{SReddy, Magstrange}. The mass modifications of pseudoscalar (P) open charm mesons $D$($D^0$, $D^+$) and $\bar{D}$($\bar{D^0}$, $D^-$) arise due to their interactions with baryons, and the scalar fields ($\sigma$, $\zeta$, and  $\delta$) in the presence of the magnetic field \cite{SReddy, Magstrange}. The interaction Lagrangian of these mesons gives rise to equations of motion of the open charm mesons, and their Fourier transforms lead to the dispersion relations for the pseudoscalar $D$ and $\bar{D}$  mesons. In the rest frame, the dispersion relation is given as 
\begin{eqnarray}
-\omega^2  + {m_{D(\bar{D})}}^2 - \Pi_{D(\bar{D})}(\omega,0) = 0,
\label{disp_relation}
\end{eqnarray}
where $m_{D(\bar{D})}$ is their mass in the vacuum. The medium effects on these mesons are incorporated in their self-energy denoted by $\Pi_{D(\bar{D})}(\omega,0)$. The expression of self energies for $D$ mesons in the magnetized strange hadronic matter is given as \cite{Magstrange}.
\begin{eqnarray}
 \Pi_{{D}} (\omega,0) &= & \frac {1}{4 f_D^2}\Big [3 (\rho_p +\rho_n)
\pm (\rho_p -\rho_n) 
+2\big(\left( \rho_{\Sigma^{+}}+ \rho_{\Sigma^{-}}\right) \pm
\left(\rho_{\Sigma^{+}}- \rho_{\Sigma^{-}}\right) \big)\nonumber\\
&+&2(\rho_{\Lambda^{0}}+\rho_{\Sigma^{0}})
+( \left( \rho_{\Xi^{0}}+ \rho_{\Xi^{-}}\right) 
\pm 
\left(\rho_{\Xi^{0}}- \rho_{\Xi^{-}}\right)) 
\Big ] \omega \nonumber \\
&+&\frac {m_D^2}{2 f_D} (\sigma ' +\sqrt 2 {\zeta_c} ' \pm \delta ')
+ \Big [- \frac {1}{f_D}
(\sigma ' +\sqrt 2 {\zeta_c} ' \pm \delta ')
+\frac {d_1}{2 f_D ^2} (\rho_p ^s +\rho_n ^s\nonumber\\
&+&\rho_{\Lambda^{0}}^s+\rho_{\Sigma^{+}}^s+\rho_{\Sigma^{0}}^s
+\rho_{\Sigma^{-}}^s
+\rho_{\Xi^{0}}^s+\rho_{\Xi^{-}}^s)
+\frac {d_2}{4 f_D ^2} \Big ((\rho _p^s +\rho_n^s)
\pm   ({\rho} _p^s -{\rho}_n^s)+\frac{1}{3}{\rho} _{\Lambda^0}^s\nonumber\\
&+&({\rho}_{\Sigma^{+}}^s +{\rho} _{\Sigma^{-}}^s)
\pm    ({\rho} _{\Sigma^{+}}^s -{\rho}_{\Sigma^{-}}^s)
+{\rho} _{\Sigma^{0}}^s \Big ) \Big ]
\omega ^2,
\label{selfenergy_D}
\end{eqnarray}
where $\pm$ refers to $D^0$ and $D^+$, respectively. For $\bar{D}$ mesons, the self-energy is given as
\begin{eqnarray}
 \Pi_{\bar{D}}(\omega, 0) &= & -\frac {1}{4 f_D^2}\Big [3 (\rho_p +\rho_n)
\pm (\rho_p -\rho_n)
 +2\big(\left( \rho_{\Sigma^{+}}+ \rho_{\Sigma^{-}}\right)\pm 
\left(\rho_{\Sigma^{+}}- \rho_{\Sigma^{-}}\right) \big)\nonumber\\
&+&2(\rho_{\Lambda^{0}}+\rho_{\Sigma^{0}})
+( \left( \rho_{\Xi^{0}}+ \rho_{\Xi^{-}}\right) 
\pm 
\left(\rho_{\Xi^{0}}- \rho_{\Xi^{-}}\right))\Big ] \omega\nonumber \\
&+&\frac {m_D^2}{2 f_D} (\sigma ' +\sqrt 2 {\zeta_c} ' \pm \delta ')
 + \Big [- \frac {1}{f_D}
(\sigma ' +\sqrt 2 {\zeta_c} ' \pm \delta ')
+\frac {d_1}{2 f_D ^2} ({\rho}_p^s +{\rho}_n^s\nonumber\\
&+&{\rho}_{\Lambda^{0}}^s+{\rho}_{\Sigma^{+}}^s
+{\rho}_{\Sigma^{0}}^s+{\rho}_{\Sigma^{-}}^s
+{\rho}_{\Xi^{0}}^s+{\rho}_{\Xi^{-}}^s)
+\frac {d_2}{4 f_D ^2} \Big (({\rho}_p^s +{\rho}_n^s)
\pm   ({\rho}_p^s -{\rho}_n^s)+\frac{1}{3}{\rho}_{\Lambda^{0}}^s
\nonumber\\
&+&({\rho} _{\Sigma^{+}}^s +{\rho} _{\Sigma^{-}}^s)
\pm ({\rho}_{\Sigma^{+}}^s -{\rho}_{\Sigma^{-}}^s)
+{\rho} _{\Sigma^{0}}^s \Big ) \Big ]
\omega ^2.
\label{selfenergy_Dbar}
\end{eqnarray}
In eq.(\ref{selfenergy_D}) and eq.(\ref{selfenergy_Dbar}),  $\sigma^\prime$ = ($\sigma - \sigma_0$), $\zeta^\prime_c$ = ($\zeta_c - \zeta_{c0}$), $\delta^\prime$ = ($\delta - \delta_0$) denotes the fluctuation of scalar fields from their vacuum values. The fluctuation $\zeta^\prime_c$ has been observed to be negligible \cite{Roder}, and its contribution to the in-medium masses of open charm mesons will be neglected in the present investigation. Here $f_D$ refers to the decay constant of $D$ mesons. The parameters $d_1$ and $d_2$ are determined by a fit of the empirical values of the Kaon-Nucleon scattering lengths \cite{Brown,Bielich,Barnes_PRC49} for I = 0, and I = 1 channels \cite{AM_PRC78_2008, AM_EURPHYS_2009}.

The dispersion relations are solved at various values of magnetic fields, strangeness fraction, and temperature to obtain the masses of these mesons. Since the mass modification of the mesons depends upon the modification of scalar density, number density, and scalar fields, the effects of baryon density, isospin asymmetry, strangeness fraction, temperature, and magnetic field get reflected on their in-medium masses. For the neutral mesons ($D^0$, $\bar{D^0}$), the in-medium effective masses are the solutions of the dispersion relations given as
\begin{eqnarray}
m^{eff}_{D^0(\bar{D^0})} = {m^*}_{D^0(\bar{D^0})} .
\label{neutral_effective_mass}
\end{eqnarray}

We assume the mass shift of the vector (V) open charm mesons $D^*({\bar{D}}^*)$ from their vacuum mass at nuclear matter saturation density, arising from the medium modification of scalar fields and baryons is similar in magnitude to the mass shift of pseudoscalar open charm mesons. Such identical mass shifts in hadronic matter are obtained for vector and pseudoscalar open charm mesons in Quark Meson coupling model \cite{Krein,Hosaka}. Hence, we have the relation $\Delta m_{D^*({\bar{D}}^*)}\equiv{m^*}_{D^*({\bar{D}}^*)}-{m}_{D^*({\bar{D}}^*)}={m^*}_{D(\bar{D})}-{m}_{D(\bar{D})} = \Delta m_{D({\bar{D}})}$ where ${m}_{D^*({\bar{D}}^*)}$ is the vacuum mass of vector open charm mesons. Moreover, the charged pseudoscalar mesons ($D^\pm$) and the longitudinal component of the charged vector open charm mesons (${D^{*\pm\parallel}}$) have additional positive mass modifications in magnetic fields through Landau quantization retaining up to the lowest Landau level and given as \cite{Magstrange,AM_SPM6, catalysis_new2, catalysis_new4}
\begin{eqnarray}
m^{eff}_{D^{\pm}} = \sqrt[]{m_{D^{\pm}}^{*2} + |eB|},\hspace{2cm}
m^{eff}_{D^{*\pm\parallel}} = \sqrt[]{m_{D^{*\pm }}^{*2} + |eB|}.
\label{charged_effective_mass}
\end{eqnarray}

In the chiral effective model, the leading order mass shift formula of the charmonium states is given as \cite{AKumar_AM_EurPhys2011, Magstrange}
\begin{equation}
\Delta m_{P,V}= \frac{1}{18}  \int dk^{2} 
\langle \vert \frac{\partial \psi (\vec k)}{\partial {\vec k}} 
\vert^{2} \rangle
\frac{k}{k^{2} / m_{c} + \epsilon} \times  \bigg ( 
\left\langle \frac{\alpha_{s}}{\pi} 
G_{\mu\nu}^{a} G^{\mu\nu a}\right\rangle -
\left\langle \frac{\alpha_{s}}{\pi} 
G_{\mu\nu}^{a} G^{\mu\nu a}\right\rangle _{0}
\bigg ), 
\label{masspsi}
\end{equation}
where 
\begin{equation}
\langle \vert \frac{\partial \psi (\vec k)}{\partial {\vec k}} 
\vert^{2} \rangle
=\frac {1}{4\pi}\int 
\vert \frac{\partial \psi (\vec k)}{\partial {\vec k}} \vert^{2}
d\Omega.
\label{integralsymbol}
\end{equation}

In the above, $\ m_{c}$ denotes the mass of the charm quark, and $\epsilon$ = $2m_c-m_{P,V}$ represents the binding energy of the  charmonium state where $m_{P,V}$ is the vacuum mass of the concerned pseudoscalar or vector charmonium state. Here, $ \langle \frac{\alpha_{s}}{\pi} 
G_{\mu\nu}^{a} G^{\mu\nu a}\rangle$ and $ \langle \frac{\alpha_{s}}{\pi} 
G_{\mu\nu}^{a} G^{\mu\nu a}\rangle_{0}$ are the expectation values of the scalar gluon condensates in the magnetized medium and in the vacuum, respectively, and $\psi(k)$ is the harmonic oscillator wave functions \cite{Amal1,AKumar_AM_EurPhys2011,Akumar_AM_PRC81_2010,Magstrange} of charmonia in the momentum space normalized as $\int \frac{d^{3} k}{(2 \pi)^{3}}|\psi(k)|^{2}=1$ .For $N_f$=3, modification of  scalar gluon condensate in the chiral effective model is given by,
\begin{eqnarray}
\bigg ( 
\left\langle \frac{\alpha_{s}}{\pi} 
G_{\mu\nu}^{a} G^{\mu\nu a}\right\rangle -
\left\langle \frac{\alpha_{s}}{\pi} 
G_{\mu\nu}^{a} G^{\mu\nu a}\right\rangle _{0}
\bigg )=\nonumber\hspace{6cm}\\\frac{8}{9}\left[(1-d)\left(\chi^{4}-\chi_{0}^{4}\right)+m_{\pi}^{2} f_{\pi} \sigma^{\prime}+\left(\sqrt{2} m_{K}^{2} f_{K}-\frac{1}{\sqrt{2}} m_{\pi}^{2} f_{\pi}\right) \zeta^{\prime}\right].
\label{massivegluoncondensate}
\end{eqnarray}
Here $\chi$ and $\chi_0 $ are the values of the dilaton field in the magnetized medium and in the vacuum, respectively. The terms proportional to $\sigma '$ = ($\sigma - \sigma_0)$ and $\zeta '$ = ($\zeta - \zeta_0$) originate due to the finite quark mass term $\sum_i m_i \bar {q_i} q_i$ in the expression of energy-momentum tensor of QCD \cite{Magstrange, Hotmagstrange}. Here, $d$ is a a parmeter related to the QCD-Beta function \cite{Magstrange}, $f_K$ is the kaon decay constant, $f_\pi$ is the pion decay constant, and $m_K$, $m_\pi$ are their respective vacuum masses. The effective mass of pseudoscalar and vector charmonium states in the medium is then given as.
\begin{equation}
{{m}^{eff}_{P,V}}=m_{P,V}+\Delta m_{P,V}
\label{effective mass_charmonium}
\end{equation}

Moreover, in the presence of the magnetic field, the coupling of the particle's spin with the magnetic field results in M1 transitions, which convert spin-1 states into spin-0 states by the emission of a photon \cite{Gubler}. Such an interaction can result in the mixing between pseudoscalar mesons and longitudinal component of vector mesons \cite{Cho, Cho1, Gubler, AM_SPM5, AM_SPM4}. The effect of spin mixing on the open charm mesons and  charmonium states are taken into account through a phenomenological effective Lagrangian interaction ${\cal L}_{PV\gamma}$ \cite{Cho, Cho1, Gubler, AM_SPM5,AM_SPM4} given as
\begin{equation}
{\cal L}_{PV\gamma}=\frac{g_{PV}}{m_{av}} e {\tilde F}_{\mu \nu}
(\partial ^\mu P) V^\nu.
\label{PVgamma}
\end{equation}
Here $g_{PV}$ is the dimensionless spin-mixing coupling parameter, $e$ is the unit electric charge, $m_{av}$ is the average masses of the pseudoscalar and the vector mesons, and ${\tilde F}_{\mu \nu}$ is the dual electromagnetic field strength tensor. The coupling parameter $g_{PV}$ is calculated from the observed value of the vacuum radiative decay width  $\Gamma(V\rightarrow P +\gamma)$ through the expression

\begin{equation}
\Gamma (V\rightarrow P \gamma)
=\frac{e^2}{12}\frac{g_{PV}^2 {p_{cm}}^3}{\pi m_{av}^2}/
\label{decay_VP}
\end{equation}
Here, $p_{cm}=(m_V^2-m_P^2)/(2m_V)$
is the magnitude of the center of mass momentum in the final state with $m_V$ and $m_P$ are the vacuum masses of the corresponding vector and pseudoscalar mesons, respectively. From the effective phenomenological Lagrangian \cite{Cho}, we obtain the masses of the pseudoscalar and the longitudinal component
of the vector mesons($V^\parallel$) after incorporating the mixing effects and are given by \cite{Cho, Cho1, Gubler, AM_SPM5, AM_SPM4}
\begin{equation}
m^{(PV)}_{P,V^{||}}=\frac{1}{2} \Bigg ( M_+^2 
+\frac{c_{PV}^2}{m_{av}^2} \mp 
\sqrt {M_-^4+\frac{2c_{PV}^2 M_+^2}{m_{av}^2} 
+\frac{c_{PV}^4}{m_{av}^4}} \Bigg),
\label{mpv_spin_Lagranian}
\end{equation}
where $M_+^2={{m}^{eff}_P}^2+{{m}^{eff}_V}^2$, $M_-^2={{m}^{eff}_P}^2-{{m}^{eff}_V}^2$ and 
$c_{PV}= g_{PV} eB$. Here ${{m}^{eff}_P}$ and ${{m}^{eff}_V}$ are the effective masses of pure states of open charm mesons and charmonia in the medium calculated in the chiral effective model. The $+$ and $-$ signs are for the vector and pseudoscalar states, respectively, indicating that the mass of the longitudinal component of the vector meson increases with the magnetic field, and that of the pseudoscalar meson drops under the effect of spin mixing. Hence spin mixing introduces a level repulsion between the mixing partners.

 For the neutral open charm mesons and charmonia, in the absence of a medium, the effect of the magnetic field is through the spin mixing effect only. However, for the charged mesons, in the absence of a medium, the effect of the magnetic field is through Landau quantization and spin mixing effect. In the hadronic medium, the modification of scalar fields, number densities, and scalar densities of baryons with respect to the changes in baryon density, isospin asymmetry, magnetic field, temperature, and strangeness fraction will also contribute to the effective masses of mesons along with spin mixing effect and Landau quantization effect.

\section{DECAY WIDTHS OF $\psi(1D)$ TO $D\bar{D}$ WITHIN $^3P_0$ model}

In this section, we describe the partial decay widths of the vector charmonium state $\psi(1D)$ to $D\bar{D}$ under strong magnetic fields using the $^3P_0$ model. In this model, a light quark-antiquark pair is created in the $^3P_0$  state, and this light quark (antiquark) combines with the heavy charm antiquark (charm quark) of the decaying charmonium state at rest, resulting in the production of the open charm ($D$, $\bar{D}$) mesons.  When spin mixing is taken into account, the general expression for the partial width of the longitudinal component of  $\psi(1D)$ is given as \cite{Friman}
\begin{eqnarray}
\Gamma_{\psi^\parallel (1D) \rightarrow  D\bar{D}} = 2\pi \frac{p_D E_D E_{\bar{D}}}{ m_{\psi^\parallel(1D)}^{(PV)}} \sum_{LS}\left|M_{LS}\right|^2, \end{eqnarray}
where $M_{LS}$ is the invariant matrix element representing the decay of the parent charmonia to  $D\bar{D}$ pairs. This matrix element will involve an overlap integral consisting of the momentum space wave functions of the parent and the daughter mesons. Here $p_D$ is  given as \cite{AM_charmdecaywidths_mag, AKumar_AM_EurPhys2011}
\begin{eqnarray}
p_D= \Bigg(\frac{ ({m_{\psi^\parallel(1D)}^{(PV)}})^2}{4}- \frac{({m_D^{(PV)}})^2 + ({m_{\bar{D}}^{(PV)}})^2} {2}+\frac{( ({m_D^{(PV)}})^2 - ({m_{\bar{D}}^{(PV)}})^2)^2}{4({m_{\psi^\parallel(1D)}^{(PV)}})^2}\Bigg) ^{1/2}.
\label{momentum}
\end{eqnarray}
Here ${m_{\psi^\parallel(1D)}^{(PV)}}$ is the in-medium mass of the longitudinal component of charmonium state and ${m_D^{(PV)}}$, ${m_{\bar{D}}^{(PV)}}$ are the medium masses of the outgoing $D$ and $\bar{D}$ mesons in the magnetic field after spin mixing. Here, $E_{D}$, $E_{\bar{D}}$ denotes the energy of outgoing $D$ and $\bar{D}$ mesons and given as
\begin{eqnarray}
E_D = \big(p_D^2 + ({m_D^{(PV)}})^2\big)^{1/2},\hspace{1cm}   E_{\bar{D}} = \big(p_D^2 + ({m_{\bar{D}}^{(PV)}})^2\big)^{1/2}.
\label{Energy_meson}
\end{eqnarray}

There are two possible decay channels for the charmonium states, through $D^0\bar{D^{0}}$ channel and through $D^+$ $D^-$ channel. The expression for partial decay widths for $\psi^\parallel(1D)$ is given as \cite{Friman, AM_charmdecaywidths_mag,AKumar_AM_EurPhys2011}
\begin{eqnarray}
  \Gamma_{\psi^\parallel(1D) \rightarrow  D\bar{D}} = \pi^{1/2} \frac{E_D E_{\bar{D}}\gamma^2}{2 {m_{\psi^\parallel(1D)}^{(PV)}}}\frac{2^{11} 5}{3^2}\Bigg(\frac{r}{1+2r^2}\Bigg)^7 x^3\Bigg(1-\frac{1+r^2}{5(1+2r^2)}x^2\Bigg)^2 \nonumber\\\times\exp \Bigg(-\frac{x^2}{2(1+2r^2)}\Bigg).
  \label{decaywidth}
 \end{eqnarray}
 
 Here $\gamma$ is the coupling strength related to the strength of the $^3P_0$ vertex and signifies the probability of creating a light quark-antiquark pair. The ratio $r=\beta/\beta_D$ is a constant for a particular charmonium state, where $\beta$ is the strength of the harmonic potential of the parent charmonium state and $\beta_D$ is the strength of the harmonic potential of the daughter $D(\bar{D})$ mesons. The scaled momentum $x$ is defined as $ x=p_D/\beta_D $. The dependence of the partial decay width of charmonia on the magnetic field, baryon density, and isospin asymmetry is encoded in the scaled momentum $x$ and the masses of charmonia and open charm mesons. When spin mixing is not considered, the masses of open charm mesons and charmonia in eq.(\ref{momentum}), eq.(\ref{Energy_meson}) and in eq.(\ref{decaywidth}) are taken to be their effective masses of pure states.

\section{RESULTS AND DISCUSSION}

\begin{figure}[htbp]
\includegraphics[height=16.5cm, width=10cm,  keepaspectratio=true]{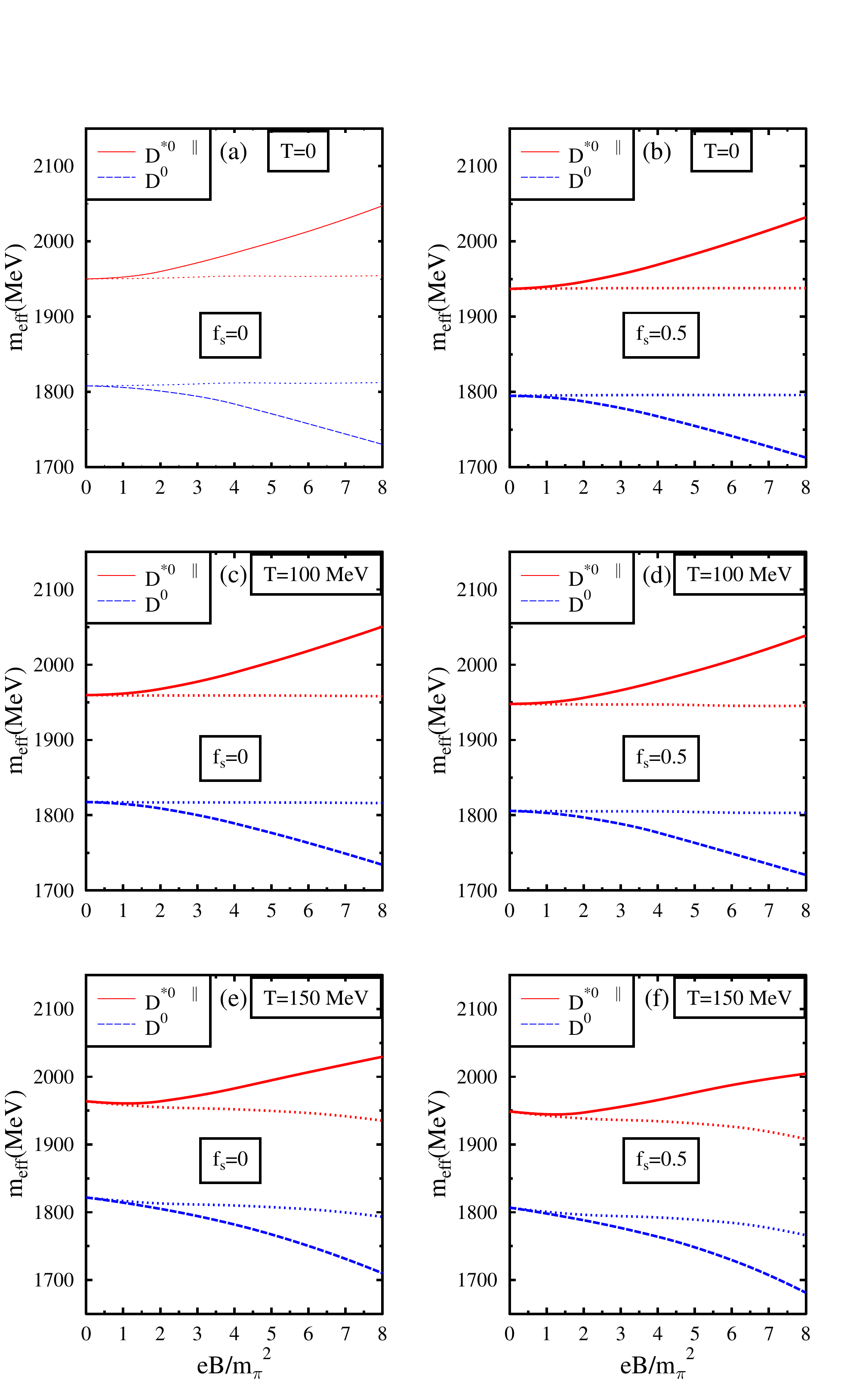}
\caption{(Color online)
The masses of pseudoscalar $D^0$ meson  and the longitudinal component of the vector $D^{*0}$ meson are plotted as functions of $eB/{m_\pi^2}$ for $\rho_B=\rho_0$ in asymmetric ($\eta$=0.5) magnetized hadronic matter for fixed values of strangeness fraction $f_s$ = 0, 0.3, 0.5 and temperature T= 0, 100 and 150 MeV. The effects of spin mixing between $D^0$ and $D^{*0\parallel}$ on their in-medium masses calculated using eq.(\ref{mpv_spin_Lagranian}) are shown and compared to the case where the mixing effects are ignored (dotted lines).}
\label{fig1}
\end{figure}

 \begin{figure}[htbp]
\includegraphics[height=16.5cm, width=10cm,  keepaspectratio=true]{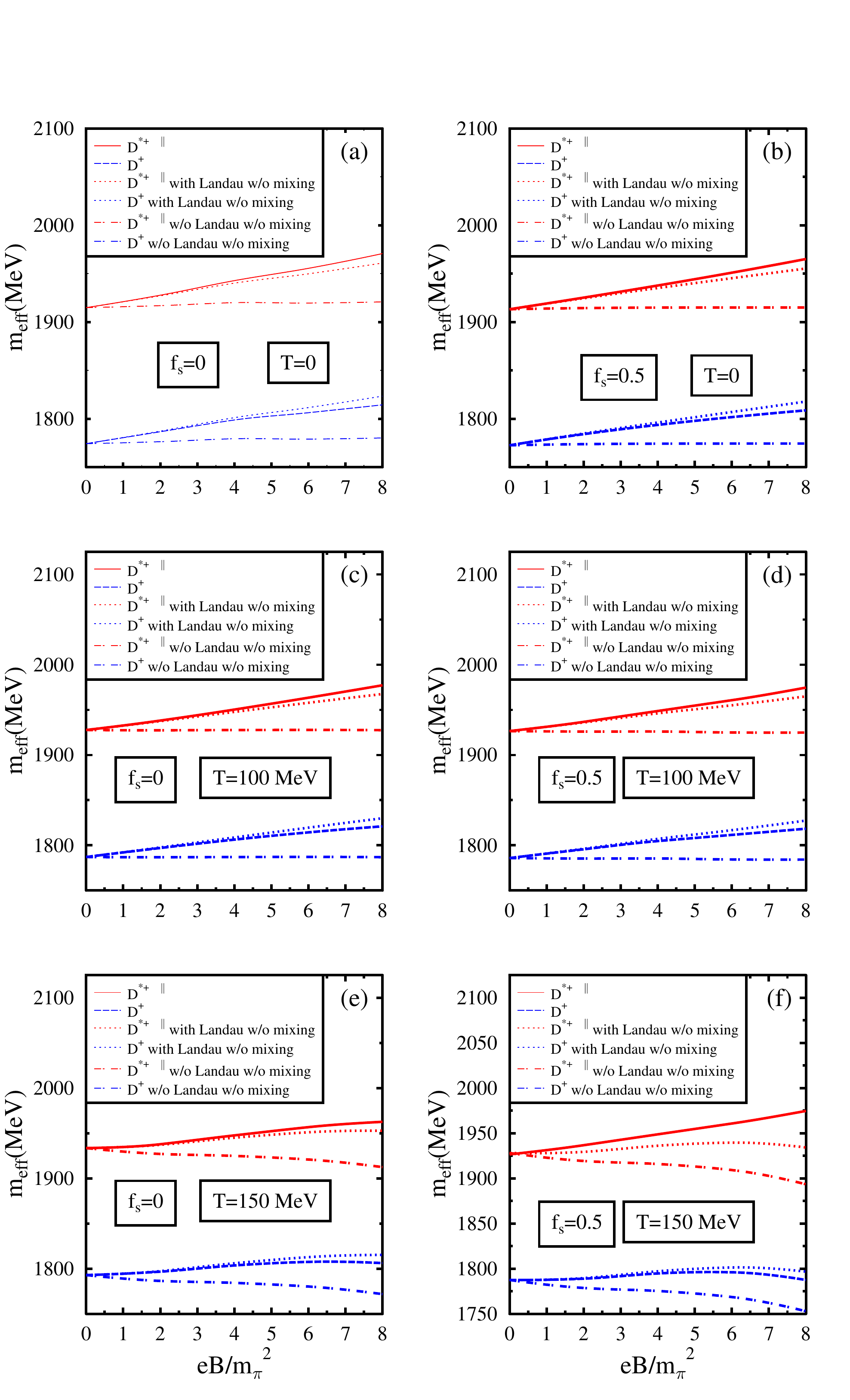}
\caption{(Color online)
The masses of pseudoscalar $D^+$ meson and the longitudinal component of the vector $D^{*+}$ meson are plotted as functions of $eB/{m_\pi^2}$ for $\rho_B=\rho_0$ in asymmetric ($\eta$=0.5) magnetized hadronic matter for fixed values of strangeness fraction $f_s$ = 0, 0.5 and temperature T= 0, 100 and 150 MeV. The effects of spin mixing between $D^+$ and $D^{*+\parallel}$, as well as Landau quantization, on their in-medium masses are shown. These plots are compared to the case where only the mixing effects are ignored (dotted lines), and both the mixing and Landau quantization effects are ignored (dashed-dotted lines).}
\label{fig2}
\end{figure}

\begin{figure}[htbp]
\includegraphics[height=16.5cm, width=10cm,  keepaspectratio=true]{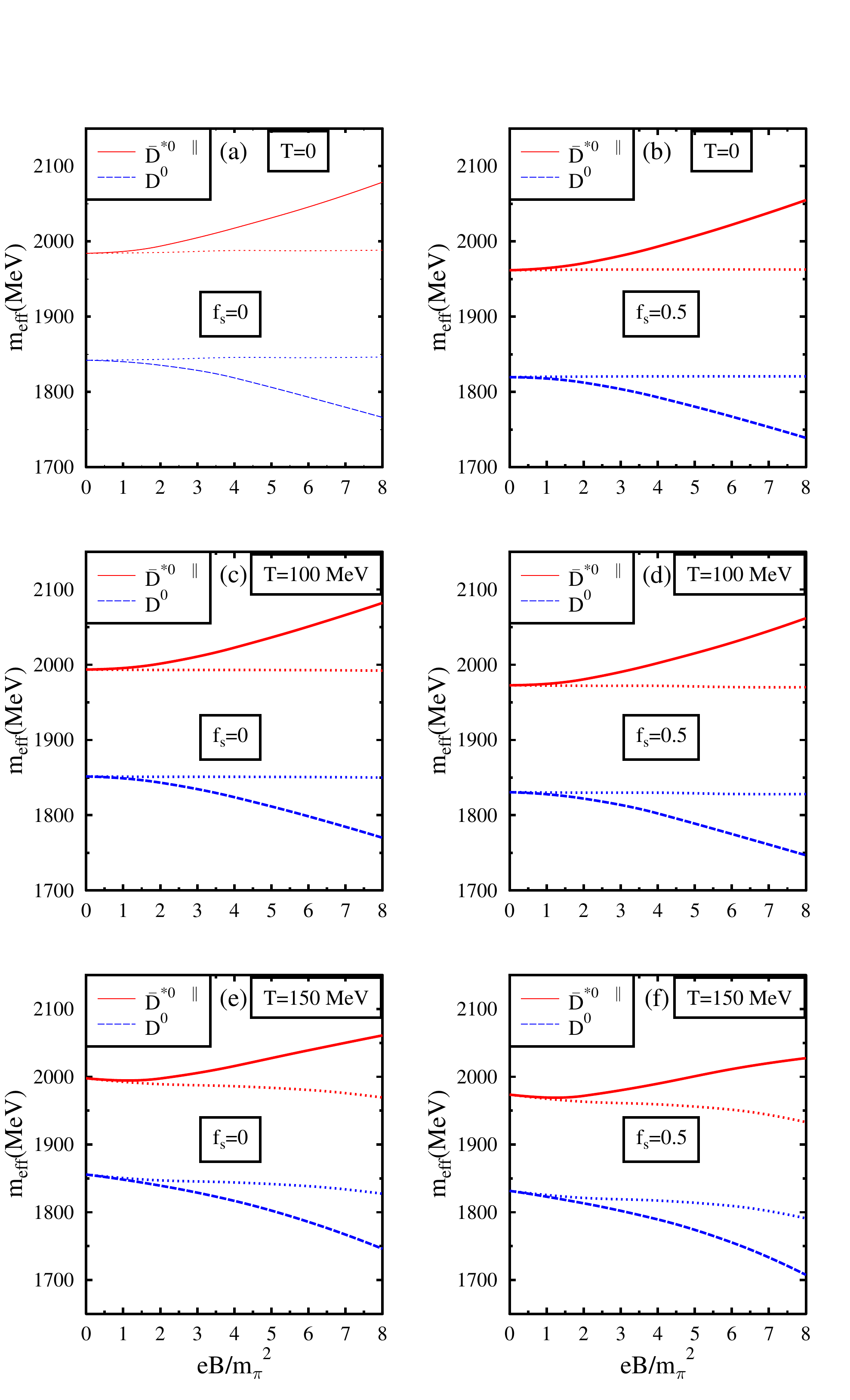}
\caption{(Color online)
The masses of pseudoscalar $\bar{D^0}$ meson  and the longitudinal component of the vector $\bar{D^{*0}}$ meson are plotted as functions of $eB/{m_\pi^2}$ for $\rho_B=\rho_0$ in asymmetric ($\eta$=0.5) magnetized hadronic matter for fixed values of strangeness fraction $f_s$ = 0, 0.5 and temperature T= 0, 100 and 150 MeV. The effects of spin mixing between $\bar{D^0}$ and ${\bar{D}}^{*0\parallel}$ on their in-medium masses are shown and compared to the case where the mixing effects are ignored (dotted lines).}
\label{fig3}
\end{figure}

\begin{figure}[htbp]
\includegraphics[height=16.5cm, width=10cm,  keepaspectratio=true]{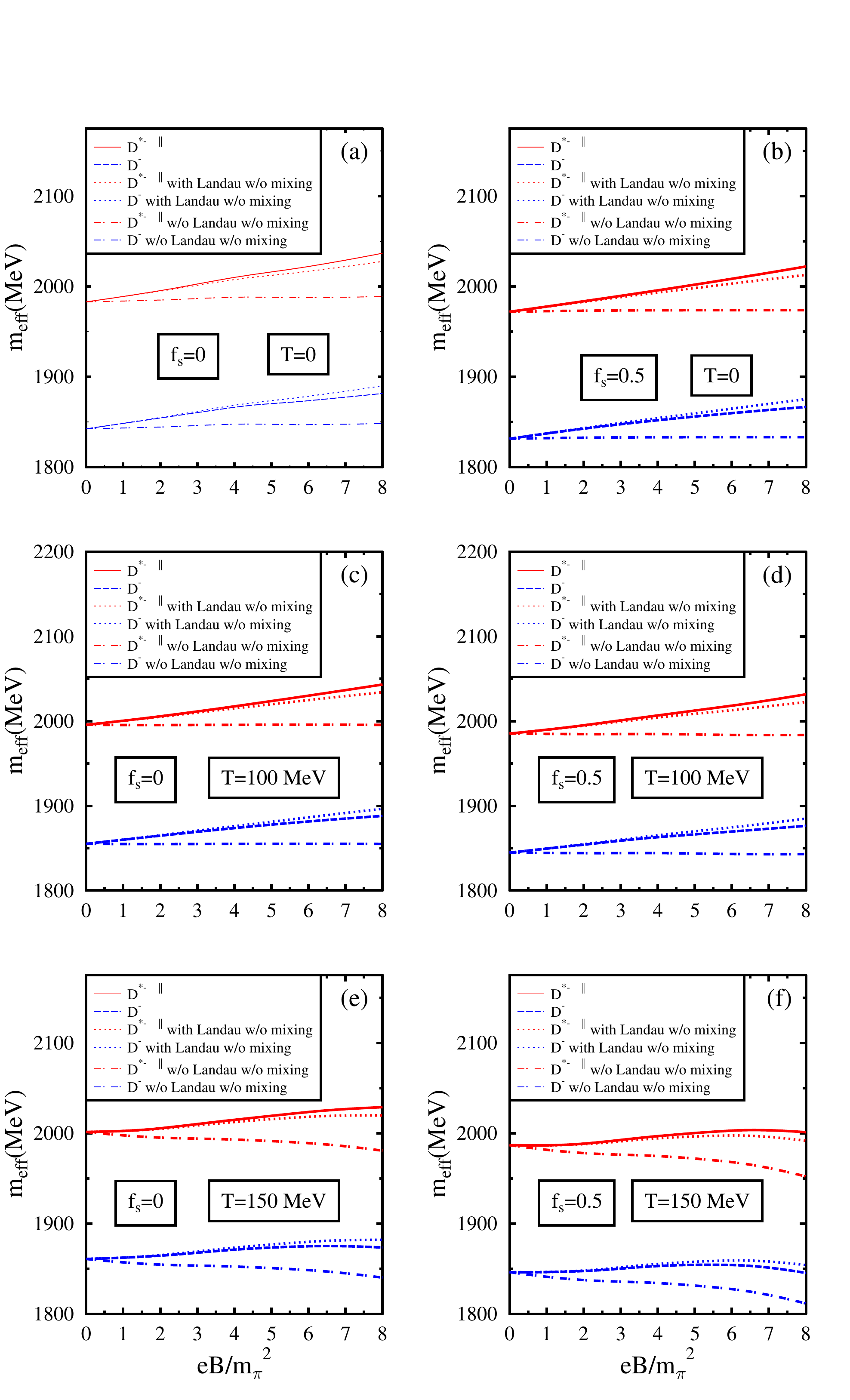}
\caption{(Color online)
The masses of pseudoscalar $D^-$ meson and the longitudinal component of the vector $D^{*-}$ meson are plotted as functions of $eB/{m_\pi^2}$ for $\rho_B=\rho_0$ in asymmetric ($\eta$=0.5) magnetized hadronic matter for fixed values of strangeness fraction $f_s$ = 0, 0.5 and temperature T= 0, 100 and 150 MeV. The effects of spin mixing between $D^-$ and $D^{*-\parallel}$, as well as Landau quantization on their in-medium masses, are shown. These plots are compared to the case where only the mixing effects are ignored (dotted lines), and both the mixing and Landau quantization effects are ignored (dashed-dotted lines).}
\label{fig4}
\end{figure}

We have investigated the masses of the pseudoscalar ($D$($D^0$, $D^+$), $\bar{D}$($\bar{D^0}$, $D^-$) and the vector ($D^*$($D^{*0}$, $D^{*+}$), ${\bar{D}}^*$(${\bar{D}}^{*0}$, $D^{*-}$) open charm mesons, as well as the pseudoscalar ($\eta_c\equiv \eta_c(1S)$, $\eta'_c\equiv \eta_c(2S)$) and the vector charmonium states ($J/\psi$, $\psi(2S)$, $\psi(1D)\equiv\psi(3770)$), in isospin asymmetric strange hadronic medium at finite temperature in the presence of strong magnetic fields. The masses of open charm mesons due to the modification of scalar fields and baryons in the hot magnetized medium are obtained by solving the dispersion relation given in eq.(\ref{disp_relation}). The mass modification of vector open charm mesons from medium modifications of scalar fields and baryons is assumed to be similar in magnitude as compared to pseudoscalar mesons. The effect of Landau quantization on charged mesons in the presence of magnetic fields is incorporated using eq.(\ref{charged_effective_mass}). The mass shift of charmonium states due to the modification of gluon condensates in the medium are obtained using eq.(\ref{masspsi}) and eq.(\ref{massivegluoncondensate}). Finally, the effects of magnetically induced spin mixing of $D^{*0}-D^{0}$, $D^{*+}-D^{+}$, $\bar{D^{*0}}-\bar{D^{0}}$ and $D^{*-}-D^{-}$ on the masses of these open charm mesons and the effects of mixing of $J/\psi-\eta_c(1S)$, $\psi(2S)-\eta_c(2S)$ and $\psi (1D)-\eta_c(2S)$  on the masses of charmonia are incorporated using eq.(\ref{mpv_spin_Lagranian}). 

The values of the various parameters of the chiral model and their fitting procedure is given in Ref. \cite{Magstrange}. The vacuum masses (in MeV) of the open charm mesons are taken to be $m_{D^{*+}}$ =$m_{D^{*-}}$= 2010.26, $m_{D^{*0}}$ = $m_{\bar{D^{*0}}}$= 2006.85, $m_{D^+}$ = $m_{D^-}$ = 1869.65 MeV and $m_{D^0}$ = $m_{\bar{D^0}}$ = 1864.8 MeV. The values of the mixing coupling parameters $g_{PV}$ $\equiv$ $g_{D^+ D^{*+}}$ is taken to be 0.9089 using eq.(\ref{decay_VP}) from the observed vacuum values of the radiative decay width of $\Gamma(D^{*+}\rightarrow D^+\gamma)$=1.33 keV \cite{AM_SPM4}. The value of $g_{D^- D^{*-}}$ is also taken to 0.9089 since $D^-$ and $D^{*-}$ being charge conjugate particles of $D^+$ and $D^{*+}$. For neutral open charm mesons, the coupling parameter $g_{D^0 D^{*0}}$ can be determined from the partial decay width $\Gamma(D^{*0}\rightarrow D^0\gamma)$ which is 35.3 percent of the total decay width of $D^{*0}$. However, the experimental value of total decay width of $D^{*0}$ is not known with sufficient accuracy. Hence the value of $g_{D^0 D^{*0}}$ is taken to be 3.426 by initially calculating the partial decay width of $\Gamma(D^{*0}\rightarrow D^0\pi^0)$ appropriately and making use of the observed branching ratio of pionic and radiative decay modes of $D^{*0}$ to calculate the partial decay width $\Gamma(D^{*0}\rightarrow D^0\gamma)$ as 19.593 keV as given in Ref.\cite{AM_SPM4}. From this value of partial radiative decay width, the value of the coupling parameter $g_{D^0 D^{*0}}$ is taken to be 3.426 in our investigation and may be compared to the value of 3.6736 given Ref.\cite{Gubler}. The value of the mixing coupling parameter $g_{\bar{D^0}\bar{D^{*0}}}$ for the decay $(\bar{D^{*0}}\rightarrow \bar{D^0}\gamma)$ is taken to be the same as the value of $g_{D^0 D^{*0}}$ due to the charge conjugation symmetry.

In Figures \ref{fig1}, \ref{fig2}, \ref{fig3}, and \ref{fig4}, the masses of the pseudoscalar and the longitudinal component of vector open charm mesons after spin mixing are plotted as a function of the magnetic field ($eB$/${m_{\pi}}^2$) in asymmetric ($\eta$ = 0.5) magnetized hadronic matter at nuclear matter saturation density ($\rho_B$= $\rho_0$ ). Each panel is plotted at a particular value of strangeness fraction ($f_{s}$) where $f_{s}$=0 corresponds to pure nuclear matter and $f_{s}$= 0.5 corresponding to strange hadronic matter and for fixed values of temperature T=0 MeV, 100 MeV, and 150 MeV. The plots are compared to the case where the spin-mixing effects are ignored, shown as dotted lines in Figures \ref{fig1} and \ref{fig3}. In Figures, \ref{fig2} and \ref{fig4}, dotted lines represent the masses without mixing effects but with only Landau quantization, and dashed-dotted lines represent when both spin mixing and Landau quantization are ignored. 

At $\rho_B$ =$\rho_0$, the masses of all open charm mesons decrease in the medium as compared to their vacuum values since the scalar densities of baryons and the fluctuations of scalar fields ($\sigma^\prime$ = ($\sigma - \sigma_0$), $\delta^\prime$ = ($\delta - \delta_0$)) increase with baryon density resulting a net attractive interaction. In the magnetized strange hadronic medium ($f_s=0.5$),  the scalar fields undergo significant modifications, and the scalar densities of hyperons also contribute to the self-energy of the open charm mesons. Hence the open charm mesons have a larger mass drop in the $f_s=0.5$ case compared to the $f_s=0$ case. In the medium, the mass degeneracy of pseudoscalar $D^+$ and $D^-$ as well as that of $D^0$ and $\bar{D^0}$ are broken due to the Weinberg-Tomozawa term in the interaction Lagrangian density \cite{AM_PRC78_2008, Magstrange}. The effect of isospin asymmetry results in a further drop in the mass of $D^+$, whereas $D^0$ experiences a positive contribution to the mass from the second term of the Weinberg-Tomozawa term. Similarly, the mass degeneracy between the vector $D^{*+}$ and $D^{*-}$ as well as that of $D^{*0}$ and ${\bar{D}}^{*0}$ will also be broken in the medium since their mass shifts due to the modification of nucleons and scalar fields are assumed to be similar to that of their pseudoscalar counterparts. The mass degeneracy of charge conjugate partners is also broken in the magnetized medium. 

When the effect of spin mixing is ignored, the masses of neutral, open charm mesons experience marginal modifications as a function of the magnetic field in isospin asymmetric ($\eta=0.5$) medium at $T$=0 MeV and $T$=100 MeV. This behavior is because the magnetic field induced modifications of scalar fields ($\sigma$ and $\delta$) and cumulative scalar densities of baryons are marginal when the temperature is not very high \cite{Hotmagstrange}. Due to the Landau quantization effect, the charged open charm mesons are subjected to additional positive mass modifications in the presence of the magnetic field. The dotted lines in Figures \ref{fig2} and  \ref{fig4} represent the combined contribution of medium effects and Landau quantization. Hence at $T$=0 MeV and $T$=100 MeV, the masses of the pseudoscalar mesons and the longitudinal component of charged vector open charm mesons increase almost linearly as a function of the magnetic field.   

However, at $T$=150 MeV, the magnitude of scalar fields significantly drops with a change in the magnetic field \cite{Hotmagstrange}, making the scalar meson exchange term more attractive with the magnetic field. Consequently, there is a significant mass drop for the neutral open charm mesons with an increase in the magnetic field at $T$=150 MeV. Even for charged mesons, at $T$=150 MeV, the positive mass modification due to Landau quantization is subdued by the negative mass shift due to the modification of scalar fields induced by the magnetic field. Hence at $T$=150 MeV, above eB= $ 6m_\pi^2$, even the masses of charged mesons drop as the magnetic field is further increased when spin mixing is ignored. In the strange hadronic medium, the effects of the magnetic field and temperature on the masses are more significant compared to the nuclear medium. The qualitative behavior of open charm mesons as a function of the magnetic field at finite temperature without spin mixing effect is similar to that of open bottom mesons investigated in Ref.\cite{Hotmagstrange}. In the medium, without spin mixing, although the individual masses of the neutral open charm meson decreases as compared to vacuum, the value of mass splitting between $D^{*0\parallel}$ and $D^{0}$ as well as   $\bar{D^{*0\parallel}}$ and $\bar{D^{0}}$ remains the same. This constant mass splitting is due to the assumption of equal mass shifts for these vector and pseudoscalar mesons from the medium modification of baryons and scalar fields.

In addition to the medium effects, when the effect of spin mixing is incorporated, the mass of the longitudinal component of neutral vector open charm mesons ($V^\parallel$) increases, and that of the neutral pseudoscalar mesons (P) drops as the magnitude of the magnetic field is increased. Hence a level of repulsion between the mixing partners is observed for neutral, open charm mesons. When spin mixing is included, at $T$=150 MeV, $D^{0}$ and $\bar{D^{0}}$ mesons experience a more significant mass drop compared to $T$=0 case due to the additional mass drop by the modification of scalar fields and scalar densities of baryons by the magnetic field. In contrast, at $T$=150 MeV, the positive mass shift experienced by $D^{*0\parallel}$ and  $\bar{D^{*0\parallel}}$ due to spin mixing is reduced due to the mass reduction by the modification of scalar fields and scalar densities of baryons by the magnetic field. The in-medium masses of open charm mesons are smaller in the magnetized strange hadronic medium than in the nuclear medium. The mixing effect is more substantial for the neutral open charm mesons due to the large value of the mixing coupling parameter compared to that of charged mesons. 

 For ${D^{*\pm}}^{\parallel}$ mesons, mass shift due to spin mixing and mass shift due to Landau quantization are both positive. Hence when the spin mixing effect is incorporated, the in-medium masses of ${D^{*\pm}}^{\parallel}$ increases further compared to the case where these effects are ignored. However, for pseudoscalar $D^{\pm}$ mesons, although the dominant Landau quantization effect results in an overall positive mass modification, the spin mixing effect contributes negatively to the mass shift and subdue the net positive mass shift above eB= $3m_\pi^2$. Hence the in-medium masses of $D^{\pm}$ decrease when spin mixing effects are incorporated compared to the case where these effects are ignored. At $T$=150 MeV, the mass drop due to the modifications of scalar fields and scalar density by the magnetic field reduces the cumulative positive mass shift experienced by vector ${D^{*\pm}}^{\parallel}$ as well as pseudoscalar $D^{\pm}$. At $T$=150 MeV, in the case of $D^{\pm}$, mass shift due to spin mixing and mass shift due to purely medium effects are in the same direction and negative. Their combined negative mass contribution can even nullify the positive mass shift due to Landau quantization at large magnetic fields. Hence in strange hadronic medium, at $T$=150 MeV, the in-medium masses of $D^{\pm}$ initially increase marginally till eB= $6m_\pi^2$ and subsequently decreases. In this case, the in-medium masses of $D^{\pm}$ at $eB= 0 m_\pi^2$ and $eB= 8m_\pi^2$ are similar in magnitude.

 Hence, the interplay of the various effects induced by the magnetic field is crucial for the charged mesons at large temperatures. Consequently, to probe the effects of the magnetic field at finite density and temperature, among the open charm mesons,  $D^0$ and $\bar{D^0}$ would be ideal candidates since they have a large negative mass shift due to spin mixing (due to large value of $g_{PV}$) as well as due to scalar field modifications. From an experimental point of view, it is also essential to quantitatively analyze the different contributions of the magnetic field. The mass splitting of ${D^{*0}}^\parallel$ and $D^{0}$ and that of ${\bar{D^{*0}}}^\parallel$ and $\bar{D^{0}}$ mesons increases with the magnetic field due to level repulsion when spin mixing is taken into account. Due to the smaller value of $g_{PV}$, the variation of mass splitting of ${D^{*\pm}}^\parallel$ and $D^{\pm}$ as a function of the magnetic field is smaller compared to that of neutral mesons.

 Including the spin mixing effect, the masses of $D^{*0\parallel}$, $D^{*+\parallel}$, $\bar{D^{*0}}^\parallel$ and  $D^{*-\parallel}$ (in MeV) in $f_s=0.5$ hadronic medium at $\rho_B=\rho_0$ and $T=0$ ,  are 1969.03 (2031.98), 1937.74 (1965.12), 1993.09 (2054.86) and 1995.81 (2022.20) respectively at eB= $ 4m_\pi^2$ ($ 8m_\pi^2$). At $T=150$ MeV, these masses in the exact order are  1965.61 (2004.57), 1938.75 (1944.24), 1989.78 (2027.51), and 1996.77(2001.10). For $f_s=0.5$, the masses of pseudoscalar $D^{0}$, $D^{+}$, $\bar{D^{0}}$ and $D^{-}$ (in MeV) at $\rho_B=\rho_0$ and $T=0$ are 1767.47 (1712.66), 1793.69 (1808.71), 1792.86 (1738.89) and 1851.96 (1866.66) respectively at eB= $4m_\pi^2$ ($ 8m_\pi^2$). At $T=150$ MeV, these masses in the same order (in MeV) become 1763.86 (1681.09), 1794.71 (1787.48), 1789.37 (1707.52), and 1852.93 (1845.25).

 \begin{figure}[htbp]
\includegraphics[height=16.5cm, width=10cm,  keepaspectratio=true]{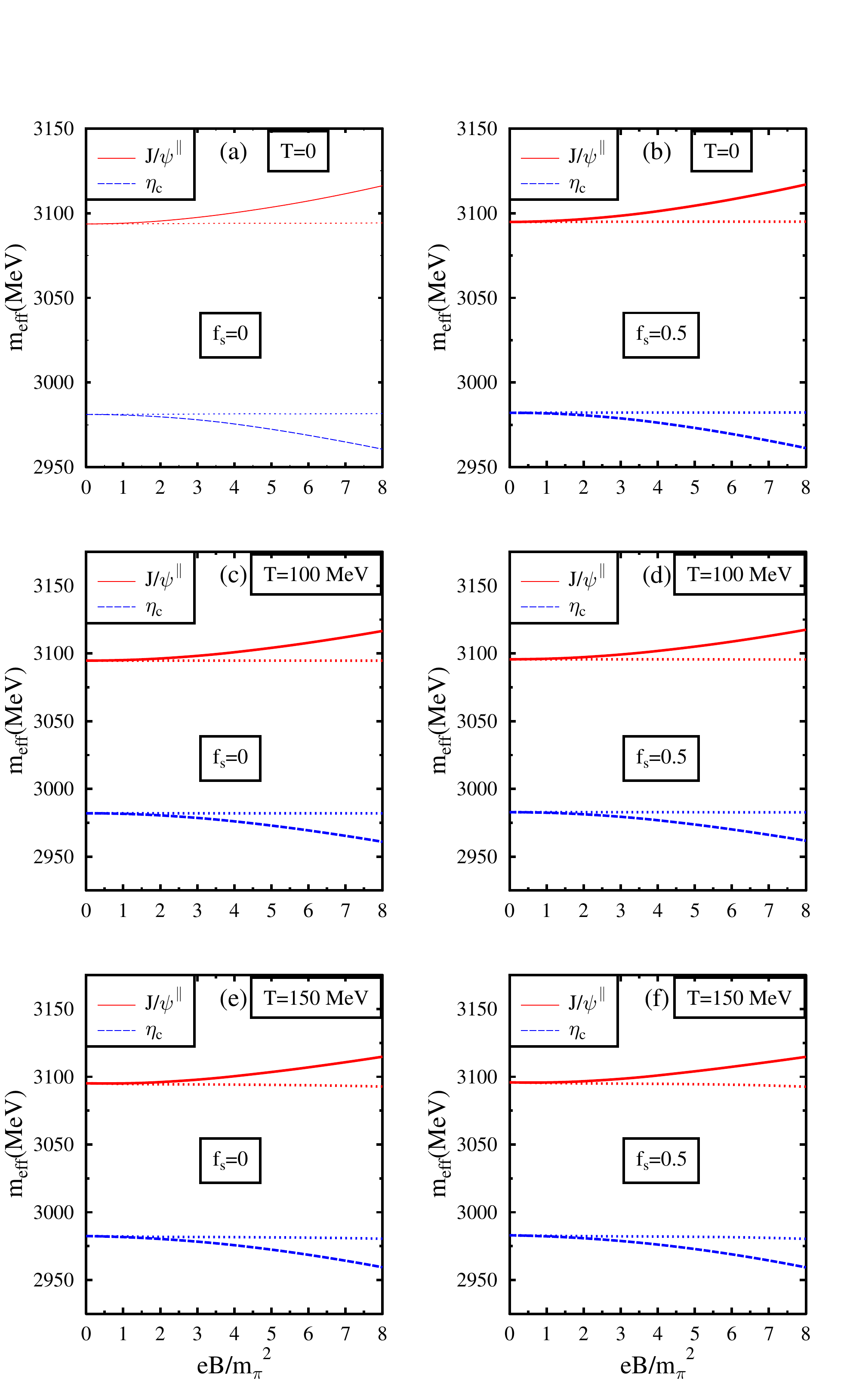}
\caption{(Color online)
The masses of the pseudoscalar charmonium state $\eta_c \equiv \eta_c (1S)$ and the longitudinal component of the vector charmonium state $J/\psi$ are plotted as functions of $eB/{m_\pi^2}$ for $\rho_B=\rho_0$ in asymmetric ($\eta$=0.5) magnetized hadronic matter for fixed values of strangeness fraction $f_s$ = 0, 0.5 and temperature T= 0, 100 and 150 MeV. The effects of spin mixing between $\eta_c$ and $J/\psi^\parallel$ on their in-medium masses are shown and compared to the case where the mixing effects are ignored (dotted lines).}
\label{fig5}
\end{figure}

\begin{figure}[htbp]
\includegraphics[height=16.5cm, width=10cm,  keepaspectratio=true]{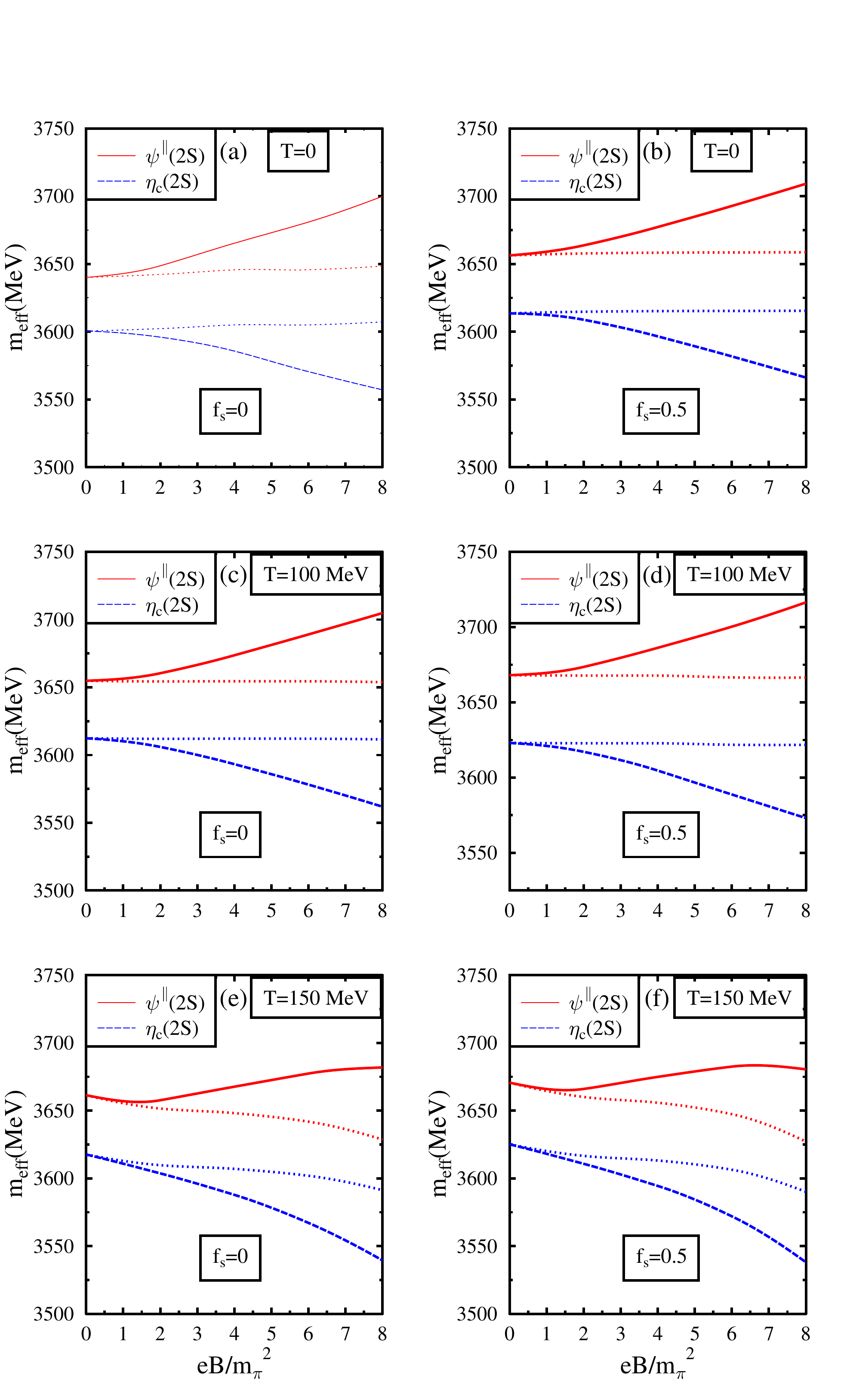}
\caption{(Color online)
The masses of the pseudoscalar charmonium state $\eta_c' \equiv \eta_c (2S)$ and the longitudinal component of the vector charmonium state $\psi(2S)\equiv \psi (3686)$ are plotted as functions of $eB/{m_\pi^2}$ for $\rho_B=\rho_0$ in asymmetric ($\eta$=0.5) magnetized hadronic matter for fixed values of strangeness fraction $f_s$ = 0, 0.5 and temperature T= 0, 100 and 150 MeV. The effects of spin mixing between $\eta_c'$ and  $\psi^\parallel(2S)$ on their in-medium masses are shown and compared to the case where the mixing effects are ignored (dotted lines).}
\label{fig6}
\end{figure}

\begin{figure}[htbp]
\includegraphics[height=16.5cm, width=10cm,  keepaspectratio=true]{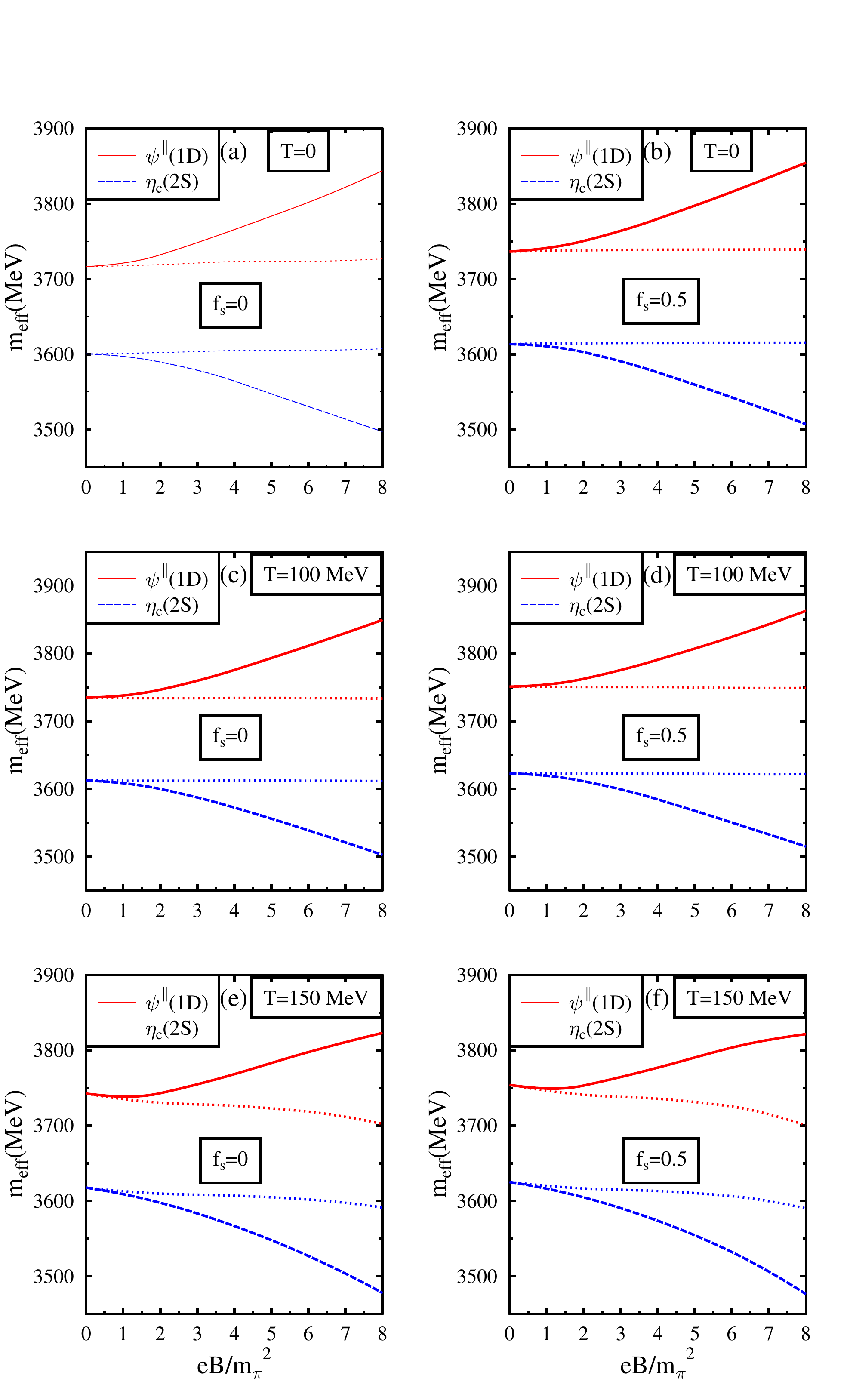}
\caption{(Color online)
The masses of the pseudoscalar charmonium state $\eta_c' \equiv \eta_c (2S)$ and the longitudinal component of the vector charmonium state $\psi(1D)\equiv  \psi(3770)$ are plotted as functions of $eB/{m_\pi^2}$ for $\rho_B=\rho_0$ in asymmetric ($\eta$=0.5) magnetized hadronic matter for fixed values of strangeness fraction $f_s$ = 0, 0.5 and temperature T= 0, 100 and 150 MeV. The effects of spin mixing between $\eta_c'$ and  $\psi^\parallel(1D)$ on their in-medium masses are shown and compared to the case where the mixing effects are ignored (dotted lines).}
\label{fig7}
\end{figure}

In Figures \ref{fig5}, \ref{fig6}, and \ref{fig7}, the masses of the longitudinal component of vector and pseudoscalar charmonium states are plotted as a function of the magnetic field ($eB$/${m_{\pi}}^2$) for different values of temperature and strangeness fraction.  The medium mass modifications of charmonia account for both the spin mixing effect as well as the modification of gluon condensates calculated using a chiral effective Lagrangian model in accord with eq.(\ref{masspsi}) and eq.(\ref{massivegluoncondensate}). The value of the parameter $\beta$, which characterizes the strength of the harmonic potential, in MeV for $J/\psi$, $\psi(2S)$ and  $\psi(1D)$  are taken to be 513, 384, and 368. They are calculated using a fit of their rms radii which are  $0.47^2$$fm^2$, $0.96^2$$ fm^2$ and $1 fm^2$ respectively \cite{CharmoniumLee,CharmoniumLee2}. For the $\eta_c$ and ${\eta_c}^\prime$  states, the values of $\beta$ in MeV are taken to be 535  and 394.6 by linear extrapolation of the vacuum mass versus $\beta$ graph of the charmonium states $J/\psi$ and $\psi(2S)$ \cite{AM_SPM5}. The values of the spin mixing coupling parameters $g_{PV}$ $\equiv$ $g_{\eta_c J/\psi}$, $g_{\eta_c^\prime \psi(2S)}$, $g_{\eta_c^\prime \psi(1D)}$ are taken to be 2.094, 3.184, 7.657 from the observed vacuum values of their radiative decay width of $\Gamma(J/\psi\rightarrow \eta_c\gamma)$=92.9 keV, $\Gamma(\psi(2S)\rightarrow {\eta_c}^\prime \gamma)$=0.2058 keV, $\Gamma(\psi(1D)\rightarrow {\eta_c}^\prime\gamma)$=24.48 keV respectively \cite{AM_SPM5}.

 Without the spin mixing effect, all the charmonium states experience a negative mass shift at finite baryon density compared to their vacuum masses. This behavior is due to the reduction in the value of $\chi$ from its vacuum value of 409.77 MeV, making the contribution of the dominant term proportional to $\chi^4-{\chi_0}^4$  in eq.(\ref{masspsi}) to be negative. In this investigation, we have also considered the effect of the quark mass term in the modification of gluon condensates through the terms proportional to $\sigma '(=\sigma-\sigma_0)$ and $\zeta '(=\zeta-\zeta_0)$ in the eq.(\ref{massivegluoncondensate}). These terms, being positive, reduce the net magnitude of the negative mass shift of charmonium states in the medium compared to the case where these terms are neglected. Since  $\sigma$ and $\zeta$ undergo significant modifications in strange hadronic medium, the quark mass term results in smaller mass shift (larger mass) for charmonium states in $f_s=0.5$ case as compared to $f_s=0$ case at $\rho_B=\rho_0$ \cite{Magstrange}. This tendency is in contrast to open charm mesons whose in-medium masses are smaller in the strange hadronic medium than in the nuclear medium. The excited charmonium states undergo more significant mass shifts compared to $J/\psi$ and $\eta(1S)$ in the medium since the momentum space integral (eq.(\ref{masspsi})) calculated for the excited state  amplifies the medium dependence of the mass shift \cite{Magstrange, Hotmagstrange}.

 The dilaton field $\chi$ increases marginally as a function of the magnetic field at $T$=0 MeV. Hence at T=0, the masses of charmonium states also increase marginally with magnetic field in $f_s=0$ case. Although the magnetic field induced mass shifts of $J/\psi^\parallel$ and $\eta(1S)$ are marginal, this is more apparent in panel (a) of the plots given in fig \ref{fig6} and fig \ref{fig7}. For $f_s=0$ case, at $T$=0 MeV,  without including the spin mixing effects, the mass of $J/\psi$, $\psi (2S)$ and $\psi (1D)$  (in MeV) are modified to 3094.02 (3094.23), 3645.68 (3648.48), 3723.31 (3726.75)  respectively at $\rho_B$ = $\rho_0$ and eB= $4m_\pi^2$ ($8m_\pi^2$) compared to their vacuum masses of 3097, 3686 and 3773 MeV. Under the same conditions, the mass of $\eta_c$  and ${\eta_c}^\prime$ (in MeV) drops to 2981.47 (2981.64) and 3605.00 (3607.25) respectively at eB= $4m_\pi^2$ ($8m_\pi^2$) compared to their vacuum masses of 2983.9 and 3637.5 MeV. 

 However, in $f_s=0.5$ case, the positive mass contribution of terms proportional $\sigma '$ and $\zeta '$ increases as a function of the magnetic field and opposes the negative mass contribution term proportional to $\chi^4-{\chi_0}^4$. Hence at T=0,  due to the interplay of these terms, the effect of the magnetic field on the mass shifts of charmonium states is smaller in the strange hadronic medium than in the nuclear medium. Also, at $T$=100 MeV, the magnetic field's effect is insignificant without the spin mixing effect. However, at $T$=150 MeV, the magnitude of $\chi$, as well as $\sigma$ and $\zeta$, drops significantly with an increase in the magnetic field. As a consequence, the contribution of the dominant term proportional to $\chi^4-{\chi_0}^4$ enhances with the magnetic field at $T$=150 MeV resulting in a larger negative mass shift for charmonium states. The behavior of charmonium states in hot magnetized strange hadronic medium without incorporating spin mixing effect is similar to that of upsilon states investigated in Ref.\cite{Hotmagstrange} 

 When the spin mixing effects are taken into account, similar to the neutral, open charm mesons, the mass of the longitudinal component of vector charmonium states ($V^\parallel$) increases, and that of the pseudoscalar charmonium states (P) drops as the magnitude of the magnetic field is increased. The magnitude of mass shift of charmonium states purely due to spin mixing is observed to be more at finite density compared to zero density case \cite{AM_SPM5}. This behavior is because ${m_V}^{eff}-{m_P}^{eff}$ is smaller at finite density compared to vacuum values, and the mass shift due to spin mixing is inversely proportional to the mass difference of the unmixed vector and pseudoscalar states at leading order \cite{AM_SPM5}. The mixing between $\psi(2S)$ with $\eta_c$ and between $\psi(1D)$ with $\eta_c$ are neglected in our calculation due to the larger mass difference between these states. 

Moreover, the contribution of the mass shift from spin mixing is observed to be larger at larger magnetic fields, as evident from the plots. The mass splitting (in MeV) between  ${J/\psi}^\parallel$ and $\eta_c$ mesons, ${\psi(2S)}^\parallel$ and ${\eta_c}^\prime$ mesons as well as   ${\psi(1D)}^\parallel$ and ${\eta_c}^\prime$ mesons increases with the magnetic field when spin mixing is taken into account. The large mass splitting for ${\psi(1D)}^\parallel$ and ${\eta_c}^\prime$ from the mixing effect is mainly due to the large value of mixing coupling parameter $g_{\eta_c^\prime \psi(1D)}$ compared to the value of $g_{\eta_c^\prime \psi(2S)}$ and $g_{\eta_c J/\psi}$. For $f_s=0.5$ and $T=0$ , including the spin mixing effect,  the masses of ${J/\psi}^\parallel$,  ${\psi(2S)}^\parallel$ and ${\psi(1D)}^\parallel$  at $\rho_B=\rho_0$, are (in MeV) 3101.14 (3116.94), 3677.27 (3709.22) and 3779.95 (3854.58) respectively at eB= $ 4m_\pi^2$ ($ 8m_\pi^2$). At $T=150$ MeV, these masses in the exact order are (in MeV)  3100.96 (3114.70), 3674.91 (3680.47), and 3777.13 (3821.67). Hence at $T$=150 MeV, the drop in the magnitude of $\chi$ due to the magnetic field reduces the net positive mass shift experienced by ${J/\psi}^\parallel$,  ${\psi(2S)}^\parallel$ and ${\psi(1D)}^\parallel$ due to spin mixing. In this case, the spin mixing effect for ${\psi(2S)}^\parallel$  becomes dominant above eB= $1.5m_\pi^2$. This behavior of $\chi$ as a function of the magnetic field at $T$=150 MeV also enhances the negative mass shift of pseudoscalar $\eta_c$ and ${\eta_c}^\prime$ mesons. 

\begin{figure}[htbp]
\includegraphics[height=16.5cm, width=10cm,  keepaspectratio=true]{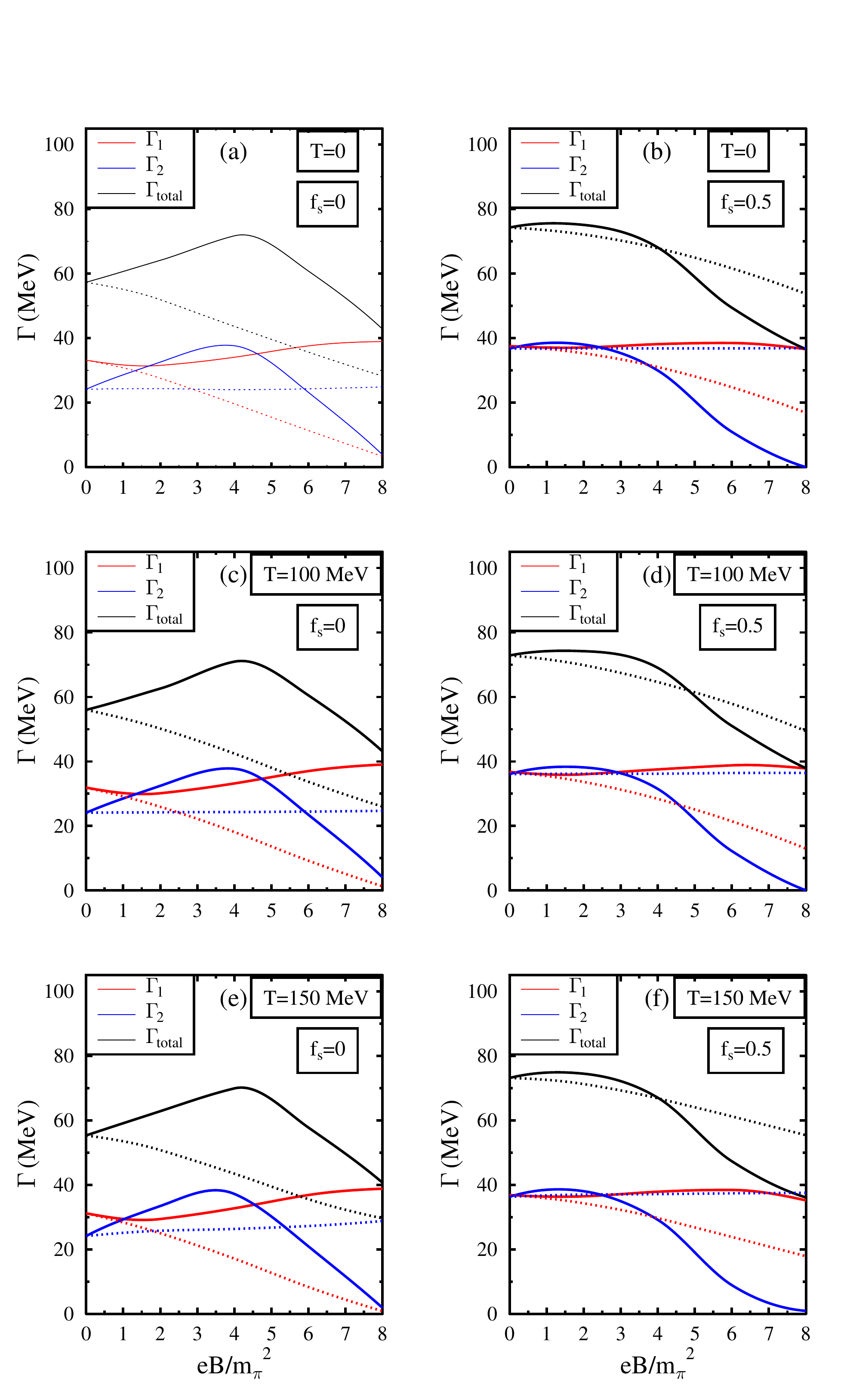}
\caption{(Color online)
Decay widths of $\psi^\parallel (1D)$ to (1) $D^+D^-$,(2) $D^0\bar {D^0}$, and the total of these two channels (1+2), are plotted as functions of of $eB/{m_\pi^2}$ for $\rho_B=\rho_0$ in asymmetric ($\eta$=0.5) magnetized hadronic matter for fixed values of strangeness fraction $f_s$ = 0, 0.5 and temperature T= 0, 100 and 150 MeV. The effects of spin mixing of both parent and daughter mesons are incorporated on the decay widths shown and compared to the case where only the mixing effects are ignored (dotted lines).} 
\label{fig8}
\end{figure}

In the present work, we have also investigated the partial decay widths of  $\psi(1D)$ to $D \bar{D}$ pairs under strong magnetic fields using the $^3P_0$ model. In this study, the value of $\beta_D$ to calculate $r$ in the expression for decay width (eq.(\ref{decaywidth})) is taken as 0.31 GeV \cite{AM_charmdecaywidths_mag}. The value of $\gamma$ is chosen to be 0.33 to reproduce the observed decay widths of $\psi(1D)$ to $D^0$ $\bar{D^{0}}$ as well to $D^+$ $D^-$ in vacuum \cite{AKumar_AM_EurPhys2011, AM_charmdecaywidths_mag}. In Figure \ref{fig8}, the partial decay widths of vector charmonium state $\psi(1D)$ to $D^+$ $D^-$  (indicated as $\Gamma_1$)  as well as to $D^0$ $\bar{D^{0}}$ (indicated as $\Gamma_2$) and the sum of these decay widths (indicated as $\Gamma_{total}$) are plotted as a function of magnetic fields in isospin asymmetric ($\eta=0.5$) hadronic matter for various values of $f_s$ and $T$. In these plots, the decay widths are shown with the effect of spin mixing on the masses of both $\psi(1D)$ and open charm mesons incorporated. The plots are compared to the case where the spin mixing effects are ignored, shown as dotted lines. In the vacuum, the partial decay width of $\psi(1D)$  to $D^+D^-$ ($\Gamma_1$) takes the value of 12.44 MeV and $\psi(1D)$  to $D^0\bar{D^{0}}$ ($\Gamma_2$) takes the value of 16.28 MeV. 

 Without considering the spin mixing effect, the masses of $\psi(1D)$ and open charm mesons drop in the medium compared to their vacuum values. At $T=0$, the masses of the $D^0$ and $\bar{D^{0}}$ mesons and $\psi(1D)$ modify negligibly with an increase in the magnetic field. Hence the value of $p_D$ and consequently $\Gamma_2$ also modify marginally in the medium with an increase in the magnetic field. Since the masses of $D^\pm$ mesons increases with the magnetic field due to Landau quantization, the value of $\Gamma_1$ reduces linearly as a function of the magnetic field. Hence at $T=0$, the value of $\Gamma_2$ is larger than $\Gamma_1$ at large magnetic fields. In the magnetized strange hadronic medium, the mass of parent meson $\psi(1D)$ is larger, and the mass of open charm mesons is smaller than their respective values in the nuclear medium. Hence the value of $\Gamma_2$ in the strange hadronic medium is observed to be larger, especially at smaller magnetic fields. 

 The spin mixing effect of $\psi(1D)$ enhances the value of its partial decay width compared to the case where these effects are not included. The effect of spin mixing on the enhancement of partial decay width is more evident in the nuclear medium than in the strange hadronic medium. When spin mixing is incorporated, the mass of ${\psi(1D)}^\parallel$ increases substantially with the magnetic field, leading to a larger value of $p_D$ for the decay. In Ref.\cite{AM_SPM5}, the partial decay width of ${\psi(1D)}$ to $D \bar{D}$ was investigated in magnetized cold nuclear medium without considering the spin mixing effect of daughter mesons. In the present investigation, when the spin mixing effect of both ${\psi(1D)}$ and open charm mesons are taken into account, the partial decay width modifies significantly, especially in the $D^0$ $\bar{D^0}$ channel due to the significant spin mixing of neutral $D$ mesons. The in-medium mass of the daughter $D^0$ and $\bar{D^0}$ drops significantly due to spin mixing, and hence the value of corresponding scaled momentum $p_D$ is larger at a fixed magnetic field, compared to the case when the spin mixing effect of open charm mesons are not taken into account. 

 For $D^+$ and $D^-$mesons, since the positive mass contribution due to Landau quantization overshoots their mass drop due to spin mixing, the values of $\Gamma_1$ modify marginally with the magnetic field and get enhanced only at large magnetic fields. This enhancement at large magnetic fields is due to the positive mass shift of the parent $\psi (1D)$ due to spin mixing effects. However, the enhancement of $\Gamma_1$ as a function of the magnetic field is suppressed in the strange hadronic medium. For $f_s=0$,  due to the increasing value of $p_D$ with the magnetic field,  $\Gamma_2$ increases initially as a function of the magnetic field up to eB=$ 3.5{m_{\pi}}^2$ to $4{m_{\pi}}^2$  and the decay width for this channel encounters a maximum at a particular value of $p_D$. Thereafter $\Gamma_2$ begins to decrease in value due to the exponential nature of the partial decay width. Hence $\Gamma_2$ encounters a maximum at intermediate values of magnetic fields in this case. For $f_s=0.5$ case, when spin mixing is incorporated, $p_D$ is larger and approaches the value of maximal point even at relatively small magnetic fields eB=$ 1.5{m_{\pi}}^2$ to $2{m_{\pi}}^2$. Hence  $\Gamma_2$ encounters a maximum at small magnetic fields, and subsequently, the value of $\Gamma_2$ decreases. Hence at large magnetic fields, due to the internal structure of wave functions of $\ psi (1D)$, the value of $\Gamma_2$ with spin mixing incorporated is small compared to the case where these effects are ignored. Hence the dominant decay of $\psi(1D)$ is through  $D^+$ $D^-$ channel at large magnetic fields above eB=$ 5{m_{\pi}}^2$ in nuclear medium and above eB=$ 3{m_{\pi}}^2$ in strange hadronic medium. The qualitative behavior total decay width ($\Gamma_{total}$) is largely determined by $\Gamma_2$ compared to $\Gamma_1$.

 The effect of the magnetic field on the individual masses of mesons is drastically different due to significant modifications of scalar fields at $T=150$ MeV, compared to the $T=0$ case. However, the effects of the magnetic field on the qualitative and quantitative nature of the decay widths at $T=150$ MeV is quite similar to that at $T=0$ case. The magnitude of decay width depends on the momentum $p_D$ (eq.(\ref{momentum})), which encodes a measure of the mass difference between $\psi(1D)$ and the combined mass of daughter mesons. Since the modification of scalar fields by the magnetic field at $T=150$ MeV contributes negatively to the mass of both parent and daughter mesons, the value of $p_D$ is similar in magnitude at $T=0$  as well as $T=150$ MeV. This results in a similar magnitude of decay widths at large temperatures. The effect of $f_s$ is more prominent than the effects of temperature since an increase in $f_s$ results in a larger mass of parent charmonium and a smaller mass of daughter mesons, thereby increasing the value of $p_D$.

The effects of magnetic field, baryon density, and temperature on the masses of open charm mesons and charmonia will have experimental consequences in ultra-relativistic heavy ion collision experiments. At LHC and RHIC, strong magnetic fields are produced. The baryon density of the produced medium in these experiments is small, but the temperature of the medium would be extremely high. To the best of our knowledge, in most previous studies investigating the effects of magnetically induced spin mixing on the properties of heavy flavor mesons, the effects of such high temperatures were not taken into account. A more baryon-dense medium with moderate temperature can be produced by reducing the collision energy. However, this would result in a decrease in the magnitude of the magnetic field produced. Strange baryons will also be present in the medium; hence, the strangeness fraction's effects are also significant. 

The magnetically induced spin mixing of open charm mesons can modify their production ratios in heavy ion collision experiments. Since the masses of pseudoscalar $D^0$ and $\bar{D^0}$ are smaller than the masses of $D^\pm$, the former mesons will be more copiously produced at large magnetic fields and high temperatures. The spin mixing of charmonia will have observational consequences on their dilepton spectra and the production of the charmonium states as well as open charm mesons in ultra-relativistic heavy ion collision experiments, e.g., at RHIC and LHC. Due to the mixing of spin-eigen states, the dileptons that should have originally arisen from vector charmonia will instead manifest from pseudoscalar charmonia resulting in anomalous decay modes $\eta_c, \eta_c^{\prime} \rightarrow l^{+} l^{-}$ \cite{AM_SPM5}. The larger masses of ${J/\psi}^\parallel$,  ${\psi(2S)}^\parallel$ and ${\psi(1D)}^\parallel$ in the magnetic field due to spin mixing suppress their production whereas the smaller mass of $\eta_c (1S)$ and $\eta_c (2S)$  due to spin mixing enhance their production. Moreover, due to the enhancement of partial decay width, the spin mixing will have observable consequences on the suppression of $\psi(1D)$ (and hence of $J/\psi$ due to feed-down effect) at small to intermediate magnetic fields. This suppression would be more significant in the strange hadronic medium than in the nuclear medium when magnetic fields are not very large.

 \section{SUMMARY}
The masses of pseudoscalar and vector open charm mesons and charmonium states in the magnetized hot hadronic medium are investigated, including the effect of magnetically induced spin mixing of these mesons. The effect of medium modifications of chiral condensates on the masses of open charm mesons and that of gluon condensates on the masses of charmonia are computed using a chiral effective model. The charged open charm mesons, $D^\pm$($D^{*\pm\parallel}$), experience additional positive mass modifications in magnetic fields through Landau quantization. At T=150 MeV, the scalar fields $\sigma$, $\zeta$, $\delta$, which mimics the chiral condensates, and  $\chi$, which mimics the gluon condensates modify significantly with the magnetic field, resulting in a more significant mass drop of all mesons when spin mixing is not incorporated. The masses of open charm mesons are smaller in the magnetized strange hadronic medium than in the nuclear medium. However, due to the presence of quark mass term, the masses of charmonium states are larger in the magnetized strange hadronic medium than in the nuclear medium.

When the spin mixing effect is incorporated, the mass of the longitudinal component of the neutral vector meson increases, and the mass of the pseudoscalar neutral mesons decreases with the magnetic field. For charged mesons, the effect of Landau quantization is observed to be dominant compared to the effect of spin mixing. At T=150 MeV, there is a strong interplay between the effects of scalar field modifications, Landau quantization, and spin mixing at large magnetic fields for charged mesons. The magnitude of spin mixing in the medium is observed to be more significant for charmonium states compared to that in the vacuum, whereas, for open charm mesons, such a medium dependence is weak. From the mass modifications of charmonium states and open charm mesons, the decay width of $\psi(1D)$ state to $D\bar{D}$  are evaluated in the $^3P_0$ model. In general, the positive mass shift of the longitudinal component of $\psi(1D)$ due to the spin mixing effect, enhances the value of its partial decay width when the magnetic field is not very large. However, the enhancement of $\Gamma_1$ is suppressed due to the positive mass shift of pseudoscalar $D^{\pm}$ mesons through Landau quantization and due to the smaller value of spin mixing coupling parameter for charged $D$ mesons. At large magnetic fields, the reduction of the mass of charged $D$ mesons due to spin mixing enhances the value of $\Gamma_1$  marginally. When the effect of spin mixing is considered, especially in the nuclear medium,  $\Gamma_2$ increases initially as a function of the magnetic field,  and thereafter $\Gamma_2$ decreases due to the exponential nature of the decay width. The effects of the strangeness fraction and the magnetic field are observed to be more significant than the effects of temperature on the decay width.

\section{ACKNOWLEDGEMENTS}
 A.J.C.S acknowledges the support towards this work from the Department of Science and Technology, Government of India, via an INSPIRE fellowship (INSPIRE Code IF170745). AM acknowledges financial support from Department of Science and Technology (DST), Government of India (project no.CRG/2018/002226).

\end{document}